\documentclass{rspublic}
\usepackage{graphicx}
\usepackage{amsmath}
\title{Statistical mechanics of collisionless relaxation in a noninteracting system}
\shorttitle{Stat. mech. of collisionless relaxation in a noninteracting system}
\author[P. de Buyl, D. Mukamel and S. Ruffo]{Pierre de Buyl$^1$, David Mukamel$^2$ and Stefano Ruffo$^{3,4}$}
\affiliation{$^1$Center for Nonlinear Phenomena and Complex Systems, Universit{\'e} Libre de Bruxelles (U.L.B.), Code Postal 231, 
Campus Plaine, B-1050 Brussels, Belgium \\
$^2$Department of Physics of Complex Systems, The Weizmann Institute of Science, Rehovot 76100, Israel\\
$^3$Dipartimento di Energetica ``Sergio Stecco'', Universit{\'a} di Firenze, and INFN, via S. Marta 3, 50139 Firenze, Italia\\
$^4$Laboratoire de Physique, UMR-CNRS 5672, ENS Lyon, 46 All{\'e}e d'Italie, 69364 Lyon cedex 07, France}

\begin{document}
\maketitle
\begin{abstract}{HMF model, Vlasov equation, Quasi Stationary States}
We introduce a model of uncoupled pendula, which mimics the
dynamical behavior of the Hamiltonian Mean Field (HMF) model. This
model has become a paradigm for long-range interactions, like Coulomb or dipolar
forces. As in the HMF model, this simplified integrable model is found to 
obey the Vlasov equation and to exhibit Quasi Stationary States (QSS), 
which arise after a ``collisionless" relaxation process. Both the 
magnetization and the single particle
distribution function in these QSS can be predicted using
Lynden-Bell's theory. The existence of an extra conserved quantity
for this model, the energy distribution function, allows us to
understand the origin of some discrepancies of the theory with
numerical experiments. It also suggests us an improvement of
Lynden-Bell's theory, which we fully implement for the zero field
case.
\end{abstract}

\def\H{\mathcal{H}}
\def\UHMF{U_\textrm{\tiny HMF}}
\def\Dth{\Delta\theta}
\def\Dp{\Delta p}
\def\etal{{\it et~al.}~}

\section{Introduction}
\label{sec:intro}

The long-range character of the interaction is responsible for
unusual properties in the thermodynamic and dynamical behaviour in a
number of physical situations (Dauxois \etal 2002, Campa \etal 2008,
Campa \etal 2009, Dauxois \etal 2009). Systems falling in this class
are self-gravitating systems, two-dimensional hydrodynamics, dipolar
interacting systems, unscreened plasmas, etc..

We will concentrate in this paper on the study of {\it kinetic 
equations}, which allow one to extrapolate from the single particle to
the collective behavior of a physical systems. 

Such equations appear also in the study of quantum transport in 
miniaturised semiconductor devices and nanoscale objects. Indeed, a 
basic model in this field is provided by the Wigner-Poisson equation 
(Manfredi, 2001), which recasts quantum dynamics in the classical 
phase space. The mathematical structure of the Wigner-Poisson system 
is analogous to that of the Vlasov equation, whose study will be the 
main topic of this paper.

Two approaches exist for the derivation of kinetic equations,
which describe the time evolution of the single particle reduced
distribution function $f(\mathbf{r},\mathbf{v},t)$. One can begin
with the Liouville equation for the full phase space distribution
function (Balescu, 1997). Going through the BBGKY hierarchy and using a
perturbative expansion, one can derive the Vlasov, Landau and Lenard-Balescu
equations when the perturbative parameter is the strength of the
potential. Alternatively, one obtains the Boltzmann equation if one uses density as
a perturbation parameter.

A second, more straightforward, approach begins instead with the
singular {\it empirical measure} in the 6-dimensional single
particle phase space and allows one to derive the Vlasov and
Lenard-Balescu equations using as perturbation parameter $1/N$, $N$
being the number of particles (Nicholson, 1992). Rigorous
mathematical approaches, which began with (Braun \etal 1977,
Neunzert 1984), are based on this second approach.

This second approach also emphasizes the role of the mean-field
potential, which is crucial when the interaction is long-range. In
the $N \to \infty$ limit, the Vlasov equation becomes exact and the
evolution towards Boltzmann-Gibbs equilibrium is hindered: the
entropy is constant in time. Evolution towards equilibrium appears
only if one considers $1/N$ corrections, which leads to ``true"
kinetic equations, see (Chavanis, 2008) for a review.

Nevertheless, already the Vlasov mean-field term induces an
evolution
of the single particle distribution function. It was originally
observed by (Henon, 1964) that this evolution produces a sort of
relaxation and that it preserves some memory of the initial state.
Following this remark, (Lynden-Bell, 1967) proposed the concept of
``collisionless" or ``violent" relaxation, which leads the system
towards a Vlasov ``equilibrium". This relaxation occurs in a time of
order $O(N^0)$, whereas later steps of relaxation of the finite $N$
system depend on $N$.

A model which has recently served as a test ground of these theories
is the Hamiltonian Mean Field (HMF) model (Antoni \& Ruffo, 1995),
which describes a system of rotators with all-to-all coupling. 
The model has been originally introduced with the aim of describing
collective phenomena in wave-particle systems of relevance for plasma
physics (Elskens \& Escande, 2002). Among the applications of interest
for this issue, let us quote the one to magnetic layered structures
(Campa \etal 2007, Dauxois \etal 2010).

For the HMF model the single particle reduced distribution function
$f(\theta,p,t)$ depends on an angle and on the angular momentum. The
Vlasov equation is exact for this model in the $N \to \infty$ limit,
as was early realized by (Messer \& Spohn, 1982). Moreover, the
$1/N$ correction to the Vlasov equation, the Lenard-Balescu term,
vanishes for one-dimensional models like the HMF model (Bouchet
2004, Bouchet \& Dauxois, 2005). This implies that one can expect
the system to evolve to equilibrium on time scales larger than
$O(N)$. Indeed, it has been found that the system evolves towards
Quasi Stationary States (QSS) (Yamaguchi \etal, 2004), which can be
interpreted as stable stationary solutions of the Vlasov equation
and whose lifetime increases as $N^{1.7}$ for homogeneous states.
Since the lifetime of QSS diverges with $N$, we expect that a QSS
will last forever in the thermodynamic limit.

QSS have been interpreted as being states that maximize Lynden-Bell
entropy (Lynden-Bell, 1967). Quite interestingly, it has been found
theoretically, and verified numerically, that, depending on the
features of the initial state, one can relax to either homogeneous
or inhomogeneous QSS and that this different evolution can be
interpreted as a phase-transition (Chavanis 2006, Antoniazzi \etal
2007$a$,$b$).

Lynden-Bell's approach has been recently successfully applied to the
free-electron laser (Barr{\'e} \etal 2004, Curbis \etal 2007, de
Buyl \etal 2009), to the one-dimensional self-gravitating sheet
model (Yamaguchi 2008, Levin \etal 2008$b$) to two-dimensional
self-gravitating systems (Teles \etal, 2010) and to models of
non-neutral plasmas (Levin \etal 2008$a$).

The theory however predicts only regimes in which all macroscopic
quantities are exactly constant in time (stationary regimes) and in
which the distribution function is a function of the energy alone.
This limitation, originating from the very same statistical nature
of the theory, causes failures whenever oscillations, non-monotonous
energy distribution or clustering regimes are present.

Several attempts have been made to give more firm grounds to
Lynden-Bell theory through a careful analysis of the self-consistent
interaction present in the Vlasov equation and of the complex
relaxation phenomenon it is responsible for. Mostly, the analysis
has been performed in the context of 1D self-gravitating systems
(Severne \& Luwel 1986, Funato \etal 1992, Yamashiro \etal 1992,
Tsuchiya \etal 1994). More recently, a description of Vlasov
equilibria in a constant mean-field potential has been also
introduced (Pomeau, 2008).

We study in this paper the Vlasov equation for a set of uncoupled
pendula and we extend Lynden-Bell's theory to this situation. The
model we study should mimic the behavior of the HMF model when this
latter has reached a steady state. Since there is no coupling among
the pendula, the Vlasov self-consistent potential reduces to a
constant potential and collisionless relaxation takes place in the
absence of variations of the potential.

In spite of all these simplifications, many of the effects found in
the HMF model are still present and are still well reproduced by the
theory. In particular, the theory predicts the value of the
magnetization attained by the system in the QSS and the occurrence
of a phase transition from a homogeneous (zero magnetization) QSS to
an inhomogeneous (magnetized) one. For our simple integrable system,
it's also easy to understand how to improve the theory by adding
conservation laws. We here propose to add conservation of moments of
the velocity distribution in the case of homogeneous states.

The paper is organized as follows. In Section~\ref{sec:system} we
introduce the Vlasov equation for the system of uncoupled pendula.
In Section~\ref{sec:numerical} we present some numerical simulations
showing the presence of homogeneous and inhomogeneous states. The
following Section~\ref{sec:LB} is devoted to the development of
Lynden-Bell's theory, which is then successfully compared with
numerical experiments in Section~\ref{sec:comparison}.
Section~\ref{sec:addition} discusses some possible improvements of
the theory, which includes an extra conservation law and
Section~\ref{sec:conclusions} presents some conclusions.

\section{Uncoupled pendula and HMF model}
\label{sec:system}

We consider the following Hamiltonian, describing a system of $N$
uncoupled pendula,
\begin{equation}
  \label{eq:NbodyH}
  H = \sum_{j=1}^N \frac{p_j^2}{2} - \H \sum_{j=1}^N \cos \theta_j ~,
\end{equation}
where $\theta_j$ is the angle of the $j$-th pendulum and $p_j$ the
corresponding angular momentum. The external field $\H$ represents
the action of gravity.

Model (\ref{eq:NbodyH}) is intended to mimic, in the large $N$
limit, the dynamics of the HMF model (Antoni \& Ruffo, 1995), whose
Hamiltonian is
\begin{equation}
\label{eq:HMF}
  H_{HMF} = \sum_{j=1}^N \frac{p_j^2}{2} - \frac{1}{2N} \sum_{i,j = 1}^N \cos(\theta_i-\theta_j).
\end{equation}

The rationale for this comparison is that, if one introduces the
magnetization
\begin{equation}
\label{eq:magnetization}
\mathbf{m} = \frac{1}{N} \sum_{j=1}^N \left( \cos \theta_j, \sin \theta_j \right) = \left( m_x,m_y \right)=m
\left( \cos \phi, \sin \phi \right)~,
\end{equation}
one can rewrite the potential of the HMF model as $-Nm^2/2$, while
the one of model~(\ref{eq:NbodyH}) reads $-N \H m_x$. It is then
evident that the two potentials have the same form if one identifies
$\H$ with $m_x/2$ and one disregards the motion of the phase of the
magnetization, i.e. one sets $\phi=0$ ($m=m_x$), which can always be
done without loss of generality taking the rotational invariance of
Hamiltonian (\ref{eq:HMF}) into account. From now on, we will then
identify the magnetization $m$ with $m_x$. The value of $m$ detects
the ``homogeneity" of the system: $m=0$ implies that the system is
homogeneous while $m > 0$ indicates an inhomogeneous system. This
terminology corresponds to a particle representation of both system
(\ref{eq:NbodyH}) and (\ref{eq:HMF}), according to which, to each
pendulum we associate a particle moving on a circle. Then, $m=1$
means that all particles are located at $\theta=0$, and therefore
the particles are fully clustered, while when $m=0$ the particles
are homogeneously dispersed on the circle. In the following, we will
make use of $m$ to quantify the state of the system and track its
evolution in time.

We will not present in this paper results for the finite $N$ case.
We will instead consider the time evolution of the single particle
reduced distribution function $f(\theta,p,t)$, which obeys the
Vlasov equation
\begin{equation}
\label{eq:vlasov_pendula}
\frac{\partial f}{\partial t} + \{ f,h \}=0~,
\end{equation}
where $\{ \cdot,\cdot \}$ are Poisson brackets and $h$ is the single
particle Hamiltonian, which, for model (\ref{eq:NbodyH}) is given by
\begin{equation}
\label{eq:oneparticle}
h(\theta,p) = \frac{p^2}{2} - \H \cos\theta~,
\end{equation}
while for the HMF model~(\ref{eq:HMF}) is
\begin{equation}
\label{eq:oneparticleHMF}
h_{HMF}(\theta,p) = \frac{p^2}{2} - m_x[f] \cos\theta -m_y[f] \sin \theta,
\end{equation}
where
\begin{eqnarray}
m_x[f] &=& \int d \theta dp \cos \theta f(\theta,p,t) \\
m_y[f] &=& \int d \theta dp \sin \theta f(\theta,p,t)~.
\end{eqnarray}
The main difference between the two single particle Hamiltonians is
that the one of model (\ref{eq:NbodyH}) does not depend on $f$, it
is simply a function of the phase space variables. The relation
between the two models is here even more evident, it amounts to
identify $m_x[f]$ with $\H$ and to put $m_y[f]=0$ (the factor 1/2
present in the previous identification was due to the presence of
pair interactions in the potential). The absence of any dependence
on $f$ in the single particle Hamiltonian makes a big difference.
Physically, it means that each particle moves on a constant energy
trajectory in an external potential which is fixed. This does not
mean that $f$ does not evolve in time, and therefore we expect also
in this case a ``collisionless" relaxation. The questions we explore
in this paper are how well Lynden-Bell's theory represents this
relaxation and how close it is to the one of the HMF model. A
crucial aspect of the relaxation of the uncoupled pendula model is
the conservation of the following energy distribution
\begin{equation}
  \label{eq:edp}
  p(e) = \int d\theta\ dp\ f(\theta,p,0) \delta(e - h(\theta,p))~,
\end{equation}
which remains constant in time, as fixed by the initial value of
$f$. Lynden-Bell's theory does not take this extra conservation law
into account: we will try in Section~\ref{sec:addition} to implement
in the theory this aspect for homogeneous states.

\section{Numerical results: homogeneous vs. inhomogeneous states}
\label{sec:numerical}

We have solved the Vlasov equation (\ref{eq:vlasov_pendula}) for the
uncoupled pendula via a semi-Lagrangian method coupled to cubic
splines. This method is based on a representation of $f$ on a grid,
i.e. $f_{i,j} = f(\theta_i,p_j)$ at the grid points
$\{(\theta_i,p_j) ; i\in [1:N_\theta] ; j\in [1:N_p]\}$. We refer
the reader to (Sonnendr{\"u}cker \etal 1999) for the semi-Lagrangian
method, or (de Buyl, 2010) for details regarding mean-field models.
The relevant parameters for these simulations are: $N_\theta$, the
number of grid points in the $\theta$-direction, $N_p$ the number of
grid points in the $p$-direction, $\Delta t$, the time-step and
$p_{max}$, which defines the size of the simulation box in phase
space as $[-\pi:\pi]\times[-p_{max}:p_{max}]$. We use in all
simulations reported in this paper the following values:
$N_\theta=N_p=256$, $p_{max}=-p_{min}=2.5$ and $\Delta t=0.1$.

\begin{figure}[h]
  \centering
  \includegraphics[width=\linewidth]{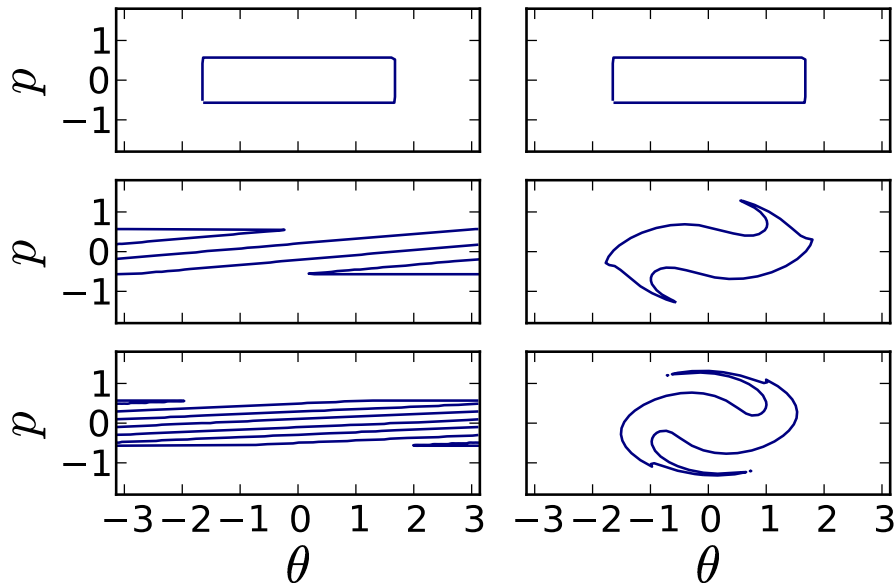}
  \caption{Evolution of the waterbag initial condition (\ref{eq:wb}) with $\Dth=1.66$ and $\Dp=0.57$. We plot the contour lines
  of the region occupied by the ``fluid level" $f_0$ in phase space with time increasing from top to bottom. The left column
  represents the case $\H=0$ and in the right column $\H = 0.75$.}
  \label{fig:phase_space_pendulum}
\end{figure}

\begin{figure}[h]
  \centering
  \includegraphics[width=0.7\linewidth]{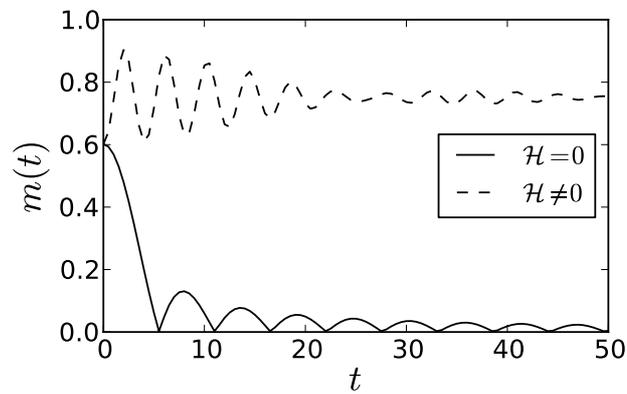}
  \caption{Magnetization $m$ vs. time for the same conditions of Fig.~\ref{fig:phase_space_pendulum}.}
  \label{fig:b_of_t}
\end{figure}

We consider a ``waterbag" initial condition: the particles, or phase
space fluid elements, occupy a rectangular region of phase space
characterized by a half width $\Dth$ in angle and $\Dp$ in momentum
\begin{equation}
  \label{eq:wb}
  \left\{\begin{array}{l l l}
      f(\theta,p,0) &=& f_0 \textrm{ if } |\theta|\leq\Delta\theta \textrm{ and } |p|\leq\Delta{}p\cr
      &=&  0 \textrm{ , else.}\cr
      & & f_0 = \left(4 \Dth \Dp\right)^{-1}
    \end{array}\right.
\end{equation}
Using this initial condition, magnetization at time zero and energy
per particle are given by $m_0=\frac{\sin\Dth}{\Dth}$ and $U=lim_{N
\to \infty} H/N=\frac{\Dp^2}{6} - \H m_0$, respectively. While the
magnetization evolves in time, the energy is a conserved quantity.
Since the Vlasov equation is a Liouville equation for $f$, the
evolution of the waterbag initial condition is such that the region
occupied by the ``fluid level" $f_0$ deforms but conserves its area
in phase space. Two qualitatively different regimes take place
depending on whether $\H$ is equal to $0$ or not. We represent the
evolution of the waterbag initial condition by the Vlasov dynamics
in Fig.~\ref{fig:phase_space_pendulum}, plotting the contour line of
the region occupied by the level $f_0$ in phase space at different
times. The time evolution of $m$ is shown in Fig.~\ref{fig:b_of_t}.

In the free-streaming case ($\H=0$), the magnetization $m$ relaxes
asymptotically to zero. Correspondingly, the initial waterbag
spreads over the whole $[-\pi:\pi]$ interval. In the $\H\neq 0$
case, a finite value of $m$ is attained at long times, with the
fluid forming a clump around $\theta=0$. This reflects the different
evolutions towards either a homogeneous state with $m=0$ or towards
an inhomogeneous state with $m \neq 0$.

\section{Lynden-Bell's theory}
\label{sec:LB}

Lynden-Bell's theory (Lynden-Bell, 1967) aims at predicting
asymptotic states of the Vlasov dynamics. The theory allows one to
compute analytically $\bar f(\theta,p)$, the {\it coarse grained}
single particle distribution function. Lynden-Bell's entropy is
derived by a combinatorial counting of the allowed states in a phase
space cell. We write here the expression of the entropy as
applicable to the waterbag initial condition in formula
(\ref{eq:wb})
\begin{equation}
  \label{eq:LBentropy}
  s(\bar f) = - \int dp\ d\theta\ \left[ \frac{\bar f}{f_0}\ln\frac{\bar f}{f_0} + \left(1-\frac{\bar f}
  {f_0}\right)\ln\left(1-\frac{\bar f}{f_0}\right)  \right] \textrm{ . }
\end{equation}
Under the constraints of ``mass", $\int f d \theta dp$, and energy
conservation, the maximization of entropy (\ref{eq:LBentropy}) for
the waterbag initial condition~(\ref{eq:wb}) leads to the following
steady state distribution
\begin{equation}
  \label{eq:LBf}
  \bar f(\theta, p) = \frac{f_0}{e^{\beta (p^2/2-\H\cos\theta) + \mu }+1}~,
\end{equation}
where $\mu$ and $\beta$ are Lagrange multipliers associated
respectively to mass and energy conservation. They can be obtained,
together with $m_{QSS}$, the magnetization in the QSS, by solving
the following set of equations
\begin{eqnarray}
  \label{eq:pendulum_LB}
  \frac{f_0 x}{\sqrt{\beta}} \int d\theta\ e^{\beta \H\cos\theta} F_0(x e^{\beta \H\cos\theta}) = 1\cr
  \frac{f_0 x}{2 \beta^{2/3}} \int d\theta\ e^{\beta \H\cos\theta} F_2(x e^{\beta \H\cos\theta}) - \H m_{QSS}  = U\cr
  \frac{f_0 x}{\sqrt{\beta}} \int d\theta\ \cos\theta\ e^{\beta \H\cos\theta} F_0(x e^{\beta \H\cos\theta}) = m_{QSS}
\end{eqnarray}
where $x=e^{-\mu}$ and the $F_n$ are defined as
\begin{equation}
  F_n(x) = \int dv\ \frac{v^n}{e^{v^2/2}+x}~.
\end{equation}
We have solved the system of equations (\ref{eq:pendulum_LB}) using
the Newton-Raphson method (Press \etal, 1996).

In the context of the uncoupled pendula model we are thus free to
choose three parameters: $\Delta\theta$, $\Delta p$ and $\H$. If we
want to compare the behavior of this model with the HMF model, we
have to reduce the free parameters to two. A physically reasonable
restriction is to impose the self-consistency condition
\begin{equation}
 \label{eq:self-c}
  m_{QSS} = \H~.
\end{equation}
Then, once $\Dth$ and $\Dp$ are given, system (\ref{eq:pendulum_LB})
allows to solve for $m_{QSS}$, $\beta$ and $\mu$ (taking into
account that $U$ can be written in terms of $\Dth$, $\Dp$ and
$m_{QSS}$), which substituted in turn into (\ref{eq:LBf}) (with
$\H=m_{QSS}$) allows us to obtain the single particle distribution
function in the QSS.

\section{Comparing with simulations}
\label{sec:comparison}

\begin{figure}[h]
  \centering
  \includegraphics[width=.8\linewidth]{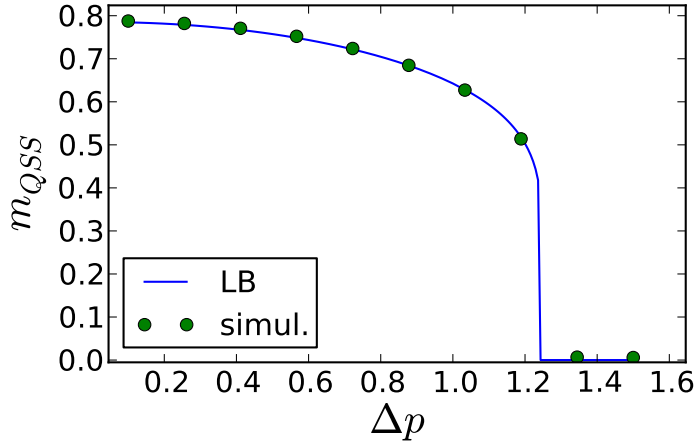}
  \caption{Comparison of the magnetization in the QSS, $m_{QSS}$, as predicted by Lynden-Bell's theory with the one
  resulting from the simulations, for a waterbag initial condition (\ref{eq:wb}) characterized by $\Dth=1.66$.}
  \label{fig:M0_060}
\end{figure}

The predicted value of $m_{QSS}$ offers a straightforward way to
compare the result of Lynden-Bell's theory with simulations. In
Fig.~\ref{fig:M0_060} we plot $m_{QSS}$ vs. $\Dp$ for $\Dth=1.66$
for both theory and numerical experiment.
The agreement in all the explored range of $\Dp$ is quite satisfactory.

A comment is however in order: if the self-consistent criterion is
fulfilled with $\H=m_{QSS}=0$, $m_{QSS}$ converges exactly to $0$ in
the simulations. This comes from the fact that the force term
$-\partial h(\theta,p)/\partial \theta$ in the Vlasov equation
vanishes and then the initial waterbag evolves under free streaming,
leading necessarily to a homogeneous distribution with respect to
$\theta$ (see the left column of Fig\ref{fig:phase_space_pendulum}).
The fact that we find an agreement also in the region where $\H \neq
0$ is less trivial.

\begin{figure}[h]
  \centering
  \includegraphics[width=.8\linewidth]{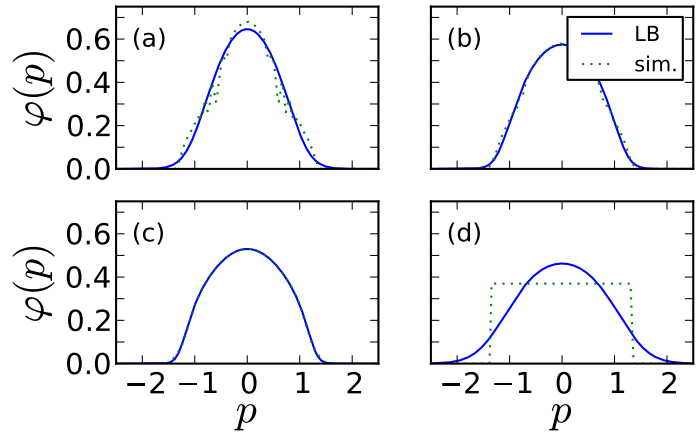}
  \caption{The marginal $\varphi(p)$ for $\Dth=1.66$ and $\Dp=0.41 (a)$,$0.72 (b)$, $1.03 (c)$ and $1.34 (d)$.}
  \label{fig:phi_comp_060}
\end{figure}

\begin{figure}[h]
  \centering
  \includegraphics[width=.8\linewidth]{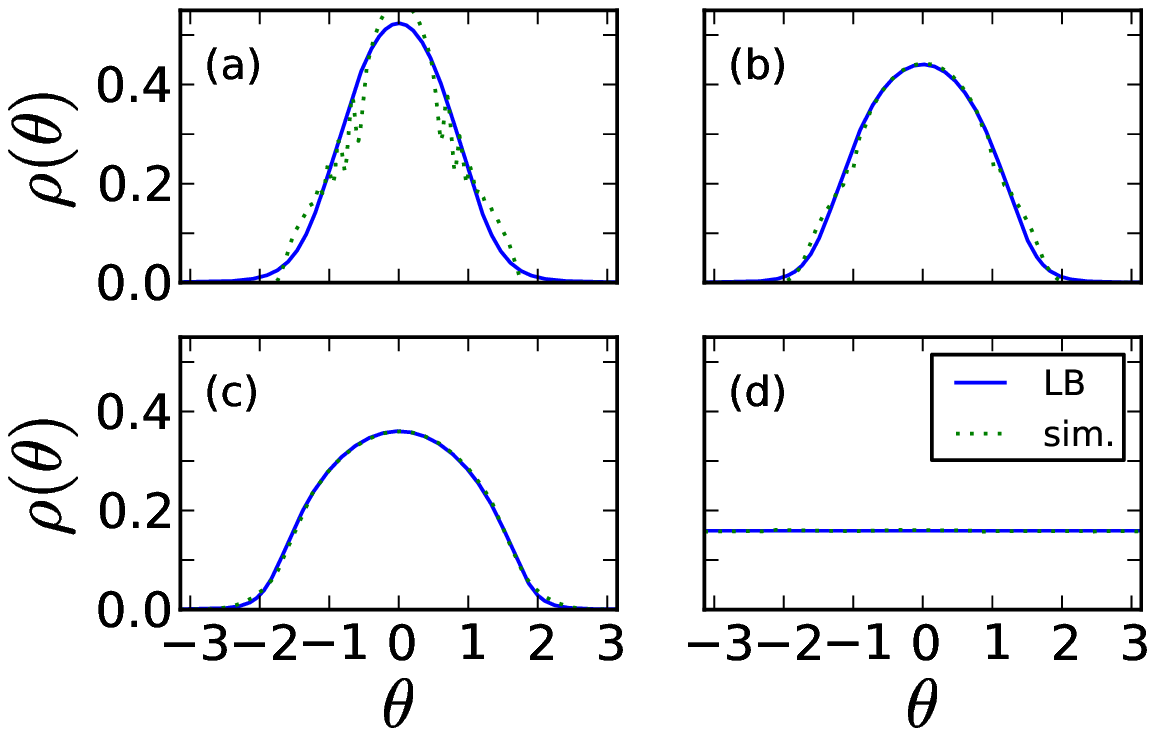}
  \caption{The marginal $\rho(\theta)$ for the same values of $\Dth$ and $\Dp$ as in Fig.~\ref{fig:phi_comp_060}}
  \label{fig:rho_comp_060}
\end{figure}

In order to explore with a finer detail the agreement of theory with
numerical experiments, we proceed to the comparison of the single
particle distribution function in the QSS. To this aim, we define
the marginal distributions
\begin{equation}
  \varphi(p) = \int d\theta\ f(\theta,p) \quad \mathrm{ ; } \quad \rho(\theta) = \int dp\ f(\theta,p) ~.
\end{equation}
In Fig.~\ref{fig:phi_comp_060} we display the marginal $\varphi(p)$
for $\Dth=1.66$ and for various values of $\Dp$ and in
Fig.~\ref{fig:rho_comp_060} the marginal $\rho(\theta)$ for the same
values of $\Dth$ and $\Dp$.

Apart from a disagreement at small scales, which is expected due to
the filamentary structure of $f$ shown in
Fig.~\ref{fig:phase_space_pendulum} (Lynden-Bell's theory provides a
prediction for the coarse-grained distribution $\bar{f}$), the
comparison in panels $(a)$, $(b)$ and $(c)$ of
Figs.~\ref{fig:phi_comp_060} and \ref{fig:rho_comp_060} is quite
good. Lynden-Bell's theory catches some basic aspects of the
spreading of the distribution along constant energy levels induced
by the Vlasov equation.  However, a major drawback of the theory
appears in panel $(d)$: while the marginal $\rho(\theta)$ is
correctly predicted, $\varphi(p)$ disagrees with numerical
simulations. This negative result is nevertheless somewhat expected,
because the time evolution with $\H=0$ is a free streaming: each
momentum level is conserved by the dynamics. This latter aspect is
not at all taken into account by the theory, which therefore fails
to reproduce $\varphi(p)$.

\begin{figure}[h]
  \centering
  \includegraphics[width=.8\linewidth]{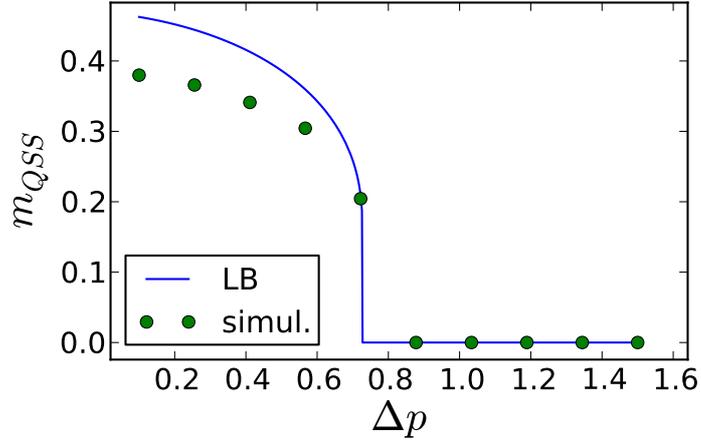}
  \caption{Comparison of the magnetization $m_{QSS}$ as predicted by Lynden-Bell's theory with the one
  resulting from the simulations. We take here $\Dth=\pi$.}
  \label{fig:M0_000}
\end{figure}

We have checked the predictions of the theory for a specific value
of $\Dth$: we cannot expect them to be reliable for all values of
$\Dth$. Indeed, the predictions of $m_{QSS}$ shown in
Fig.~\ref{fig:M0_000} for $\Dth=\pi$ are worse, although the
transition value of $\Dp$ is well reproduced.

\begin{figure}[h]
  \centering
  \includegraphics[width=.8\linewidth]{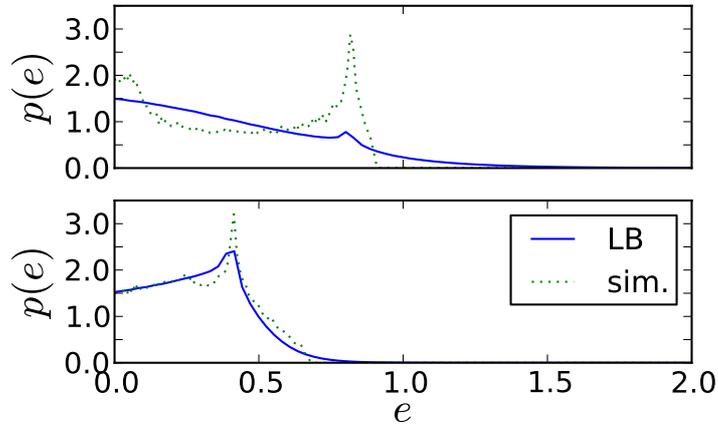}
  \caption{Energy distributions $p(e)$ given by formula (\ref{eq:edp}) for $\Dth=\pi$, $\Dp=0.41$ (top panel)
  and $\Dth=\pi$, $\Dp=0.72$ (bottom panel), computed using Lynden-Bell's theory and numerically.}
  \label{fig:p_of_e}
\end{figure}

Since the source of disagreement, as we have observed above, can be
in the failure of the theory to take into account extra conservation
laws, we compare in Fig.~\ref{fig:p_of_e}, the energy distribution
(\ref{eq:edp}) predicted by the theory with the one resulting from
the numerics (this latter being constant in time and fully
determined by the initial condition). As expected, when the value of
$m_{QSS}$ is better predicted (e.g. for $\Dth=\pi$ and $\Dp=0.72$),
the energy distribution found using Lynden-Bell's theory is closer
to the initial one.

We therefore argue that the success of Lynden-Bell's theory in the
case of uncoupled pendula originates partly from the fact that the
initial energy distribution $p(e)$ is close enough to the one
predicted by Lynden-Bell's theory itself. This may look accidental
and throw some doubts on the general validity of the theory.
However, the excellent agreement found for $\Dth=1.66$ remains a
good indication of the quality of the predictions of the theory.
Further systematic work of comparison of the theory with numerical
experiment is required in order to assess the origin of
disagreements.

\section{Adding conserved quantities in the zero field case}
\label{sec:addition}

The failure of Lynden-Bell's theory to reproduce $\varphi(p)$ for
the homogeneous case ($\H=0$) could be easily understood because of
the trivial dynamics involved. A deeper remark is that the
conservation of the energy distribution $p(e)$ reduces for $\H=0$ to
the conservation of $\varphi(p)$. We here then impose this
constraint by requiring the conservation of the even moments of
$\varphi(p)$
\begin{equation}
P_{2n}[f] = \int d\theta\ dp\ f(\theta,p) p^{2n} \quad \quad n=1,2,\dots~,
\end{equation}
which is a consequence of the conservation of each individual
momentum for the uncoupled pendula with $\H=0$. The odd moments are
all zero for symmetry reasons. The $n=0$ moment is the total ``mass"
and its conservation is already imposed in the Lynden-Bell's theory.
The $n=1$ moment is the energy and has already been considered: we
denote the corresponding theory by LB1. We restrict here to impose
the extra conservation of the $n=2$ and $n=3$ moments: the
corresponding theories will be denoted as LB2 and LB3.

Maximizing Lynden-Bell's entropy (\ref{eq:LBentropy}) leads to the following coarse grained distribution
function
\begin{equation}
\bar f(\theta,p) = \frac{f_0}{e^{\beta p^2/2 + \sigma_2 p^4 + \sigma_3 p^6+\mu}+1}~,
\end{equation}
where $\beta$ is the Lagrange multiplier conjugated to the energy,
$\sigma_2$ and $\sigma_3$ those corresponding to the $n=2$ and $n=3$
moments. The new system of equations to be solved is
\begin{subequations}
\label{eq:LBn}
  \begin{align}
  f_0\ x\ 2\pi\ G_0(x,\beta, \sigma_{2,3}) = 1 \\
  f_0\ x\ \pi\  G_2(x, \beta, \sigma_{2,3}) = U \\
  f_0\ x\ 2\pi\ G_{2n}(x, \beta, \sigma_{2,3}) = P_{2n}[f_{|t=0}] \quad \quad n=2,3
  \end{align}
\end{subequations}
where $G_{2n}(x, \beta, \sigma_{4,6}) = \int dv\ v^{2n} (e^{\beta v^2/2 + \sigma_2 v^4 + \sigma_3 v^6}+x)^{-1}$.

\begin{figure}[h]
  \centering
  \includegraphics[width=\linewidth]{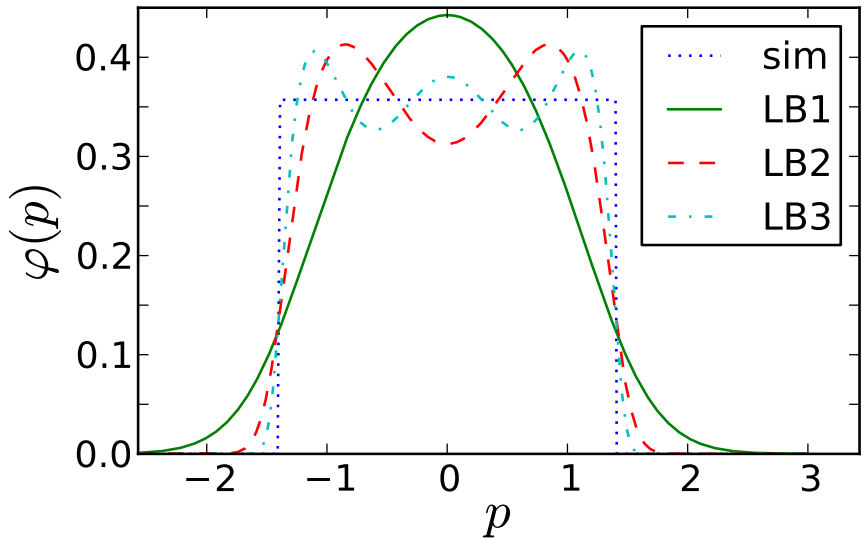}
  \caption{Comparison $\varphi(p)$ obtained from LB1, LB2 and LB3 theories (see text) with the simulation for
  $\Dp = 1.4$ and $\Dth=1.66$. For the LB3 theory the values of the Lagrange multipliers are: 
  $x = 1.29$, $\beta = 3.98$, $\sigma_2=-14.45$ and $\sigma_3 = 12.30$.}
  \label{fig:order_246}
\end{figure}

We find the result displayed in Fig.~\ref{fig:order_246}. The
presence of two humps in $\varphi(p)$ for the LB2 theory and three
humps for LB3 looks surprising, but it should be interpreted as
similar to the Gibbs phenomenon: approximating a function with a
limited set of basis functions can give rise to oscillations.
Increasing the number of conserved moments of $\varphi(p)$ can only
lead to the waterbag distribution, since it is the only one that has
the good values of $P_{2n}[f]$ for all $n$'s.

It is worth remarking that, for high values of $\beta$, the
LB1 theory is able to reproduce a step profile accurately enough, 
providing good results for the homogeneous situations considered here
(one can compare this limit with the low temperature Fermi distribution).
Our extension to the LB2 and LB3 theories allows us to obtain a better 
match in all situations where $\beta$ is smaller, as illustrated in 
Fig.~\ref{fig:order_246}. Indeed, lower values of $\Dth$ (at constant
$\Dp$) favour smaller values of $\beta$. On the other hand, one can
prove, by a direct inspection of Eqs.\ref{eq:LBn}, that, at fixed $\Dth$, 
$\beta$ scales as $\Dp^{-2}$ and $\sigma_{n}$ as $\Dp^{-2n}$. 

Let us mention that, although computationally more complex, one
could in principle obtain improved theories in the $\H \neq 0$ case
by imposing the conservation of the moments of the energy
distribution $p(e)$.
For the HMF model this approach is unfortunately not viable because $p(e)$ is not conserved.

\section{Conclusions}
\label{sec:conclusions}

With the aim of reaching a deeper understanding of the Quasi
Stationary States (QSS) observed in the HMF model, we have studied
in this paper the Vlasov equation of a set of uncoupled pendula.

These studies are relevant for understanding the extrapolation
from single particle to collective properties in physical systems
with unscreened long-range interactions, like Coulomb or dipolar. 
They will also lead to a better understanding of kinetic
equations, like the Vlasov equation, whose mathematical structure
appears in many different fields, including quantum transport at the
nanoscale with the use of the Wigner-Poisson system.

Lynden-Bell's theory applies successfully to the uncoupled pendula model. We
have found good predictions for both the magnetization and the
marginals $\varphi(p)$ and $\rho(\theta)$ of the single particle
distribution function in the QSS, although the agreement with
numerical simulations depends on the chosen initial state.

The discrepancy that we observe between the theory and the numerical
results has provided us with a hint on how to extend the theory by
including the moments of the velocity distribution for the zero
field case. This is justified by the presence of a further conserved
quantity for the uncoupled pendula case: the energy distribution
$p(e)$ in formula (\ref{eq:edp}).

Unfortunately this approach is not straightforwardly applicable to
the HMF, because the energy distribution is not conserved. However,
it has been recently realized that the steady state distribution of
the model of uncoupled pendula can be obtained analytically (de Buyl
\etal, 2010) and shows properties very similar to those of the HMF
model.

\begin{acknowledgements}
PdB would like to thank P. Gaspard for his support and acknowledges
financial support by the Belgian Federal Government
(Inter-university Attraction Pole ``Nonlinear systems, stochastic
processes, and statistical mechanics'', (2007-2011). SR thanks
UJF-Grenoble and ENS-Lyon for hospitality and financial support. DM
thanks the support of the Israel Science Foundation (ISF) and the
Minerva Foundation with funding from the Federal German Ministry for
Education and Research.

\end{acknowledgements}

\end{document}